\documentclass[aps,pre,twocolumn,longbibliography]{revtex4-2}

\usepackage{amsmath,amssymb,mathtools,bm}
\usepackage{graphicx}
\usepackage{xcolor}
\usepackage{hyperref}

\begin{document}

\title{Explicit Construction of Floquet--Bloch States from Arbitrary Solution Bases of the Hill Equation}

\author{Gregory V. Morozov}
\email{gregory.morozov@uws.ac.uk}
\affiliation{Scottish Universities Physics Alliance (SUPA), University of the West of Scotland, Paisley, United Kingdom}

\date{\today}

\begin{abstract}
For the Hill equation describing one-dimensional periodic systems, 
an explicit construction of Floquet--Bloch states from arbitrary pairs of linearly independent solutions is developed. 
Closed-form formulas map an arbitrary fundamental system to a Floquet--Bloch basis via the monodromy matrix, 
including the generic Jordan band-edge case, without reliance on canonically normalized solutions. 
An equivalent transfer-matrix formulation makes the residual representation freedom transparent 
and provides a practical framework for analytical and numerical studies of periodic systems.
\end{abstract}

\keywords{Floquet theory, Hill equation, monodromy matrix, Floquet--Bloch states, periodic differential equations}

\maketitle

\section{Introduction}

This paper develops an explicit construction of Floquet--Bloch states in one-dimensional (1D) periodic structures
from an arbitrarily chosen fundamental system of the associated Hill equation,
a second-order linear ordinary differential equation with periodic coefficients.
We restrict attention to real-valued coefficients and impose no normalization on the fundamental solutions.

The existence and properties of Floquet--Bloch states are classical results 
of Floquet theory and the spectral theory of periodic differential equations.
This work does not revisit these results.
Instead, it provides a constructive, basis-independent framework 
that renders the transformation from an arbitrary fundamental system 
to the Floquet--Bloch basis fully explicit via closed-form algebraic formulas.

As a concrete realization, we consider light propagation in layered 1D periodic dielectrics (photonic crystals),
where the periodic coefficient arises from a real-valued refractive-index profile.
From Maxwell's equations, the fields of linearly polarized light of angular frequency $\omega$
propagating along the periodic axis $z$ may be written as
\begin{equation}
\mathbf{E}(\mathbf{r}, t) = E(z) e^{-i\omega t} \hat{\mathbf{y}}, \quad
\mathbf{B}(\mathbf{r}, t)  = \frac{i}{\omega} \frac{d E(z)}{d z} e^{-i\omega t}\hat{\mathbf{x}},
\end{equation}
so that $E(z)$ satisfies
\begin{equation}
\frac{d^{2} E(z)}{d z^{2}} + k^{2} n^{2}(z) E(z) = 0, \quad n(z)=n(z+d), 
\end{equation}
where $k=\omega/c$.
For each $k$, Eq.~(2) admits infinitely many fundamental systems of solutions, among which the Floquet--Bloch system is distinguished.

According to Floquet theory \cite{McLachlan1947,Stoker1950,MagnusWinkler1966,Eastham1975,YakubovichStarzhinskii1975,Fedoryuk1985},
this system contains at least one non-trivial solution $F_1(z)$ satisfying
\begin{equation}
F_1(z+d)=\rho_1 F_1(z),
\end{equation}
where $\rho_1$ is a Floquet multiplier.
A second linearly independent solution $F_2(z)$ satisfies either
\begin{equation}
F_2(z+d)=\rho_2 F_2(z), \quad \rho_1\rho_2=1,
\end{equation}
or, in the degenerate case $\rho_1 = \rho_2 \equiv \rho$, obeys
\begin{equation}
F_2(z+d)=\rho F_2(z)+\theta F_1(z), \quad \theta\neq 0.
\end{equation}
Solutions satisfying $F(z+d)=\rho F(z)$ are called Floquet--Bloch waves. 
They admit the representation
\begin{equation}
F_{1,2}(z) = P_{1,2}(z)e^{\pm i\mu z},  \quad   \rho_{1,2} = e^{\pm i\mu d},
\end{equation}
where $P_{1,2}(z) = P_{1,2}(z+d)$ are periodic functions and $\mu$ is the Bloch wavenumber.
A solution satisfying Eq.~(5) is called a hybrid (Jordan) Floquet mode.
It admits the representation
\begin{equation}
F_2(z)=\left[\theta\frac{z}{\rho d}P_1(z)+P_2(z)\right]e^{i\mu z},
\quad \theta \neq 0.
\end{equation}
Throughout this paper we fix $\theta=1$.

Conventionally, the Floquet--Bloch system is constructed 
from the canonically normalized solutions $u(z)$ and $v(z)$ of Eq.~(2),
\begin{equation}
u(0)=1, \quad u^{\prime}(0)=0, \quad v(0)=0, \quad v^{\prime}(0)=1,
\end{equation}
see Refs.~\cite{McLachlan1947,Stoker1950,MagnusWinkler1966,Eastham1975,Cottey1972,Nusinsky2006,Morozov2011-3,Morozov2017-1}.
Standard treatments fix such a canonical normalization at the outset,
leaving the transformation from an arbitrary fundamental system largely implicit.
In many physical and numerical contexts, for example,
transfer-matrix descriptions of layered media or traveling-wave scattering formulations,
solutions arise naturally in non-canonical bases.
It is therefore advantageous to have closed-form formulas
that map the available fundamental matrix directly to a Floquet--Bloch fundamental system
without re-solving the differential equation with prescribed canonical initial data.

Near spectral transition points where the Floquet multipliers coalesce,
the second Floquet--Bloch state generally acquires a hybrid (Jordan) structure.
Treating this case directly in terms of the given fundamental system
yields an explicit and representation-transparent construction
of the unique (up to scaling) Floquet--Bloch wave
together with its associated hybrid (Jordan) mode.

Recent developments in periodically driven (``Floquet-engineered") systems
have renewed interest in Floquet--Bloch states and their associated band
structures~\cite{Bukov2015,Eckardt2017}. In such systems, 
the available solutions are likewise not given in the canonically normalized form. 
This occurs, for example, in studies of driven lattice models, Floquet band
structures, and related topological and dynamical-symmetry settings
~\cite{Ikeda2018,Asboth2014,Minguzzi2022,Wang2023,Coppini2024}. 
This motivates a formulation in which Floquet--Bloch states can be constructed
explicitly from an arbitrary fundamental system of solutions.

In this work, we develop the constructive formulation outlined above. 
Starting from an arbitrary fundamental system, 
we express the Floquet--Bloch construction in terms of the monodromy matrix 
and its canonical (diagonal or Jordan) forms. 
Each choice of fundamental system generates a monodromy matrix, 
and different choices produce conjugate matrices associated with the same differential equation.
Their spectra are therefore identical and coincide with the spectrum of the one-period transfer matrix. 
Consequently, Floquet--Bloch states may equivalently be characterized as eigenstates of the transfer matrix for one period. 
The analysis distinguishes intrinsic spectral features (multipliers, Bloch wavenumber, spectral classification) 
from basis-dependent ones (normalization and hybrid-mode mixing).

In summary, the main contributions of this work are as follows:
\begin{enumerate}
\item 
We derive explicit closed-form transformation formulas that map an
arbitrary fundamental matrix to the corresponding Floquet--Bloch
fundamental system via reduction of the monodromy matrix to its
canonical (diagonal or Jordan) form.
\item 
We provide a complete classification of the residual representation freedom
of Floquet--Bloch states induced by the choice of a fundamental system:
independent rescalings in the nondegenerate case, and
upper-triangular mixing of the hybrid mode with the periodic
(or antiperiodic) Floquet--Bloch wave in the degenerate (Jordan) case.
\item 
We reformulate the construction directly in terms of the one-period transfer matrix,
an intrinsic propagation operator of the differential equation 
whose eigenvectors correspond to Floquet--Bloch states. 
This formulation allows Floquet--Bloch states to be obtained directly from the transfer matrix,
without introducing canonically normalized solutions or constructing monodromy matrices explicitly. 
As a result, the method becomes directly compatible 
with standard transfer-matrix approaches in the physics of layered media.
\item 
We organize the results into a systematic step-by-step procedure
that translates the abstract Floquet structure into a practical
workflow for analytical and numerical implementations.
\end{enumerate}

Taken together, these results complement the classical spectral theory
by rendering the basis transformation underlying Floquet--Bloch states
fully explicit and operational.

Although the formalism applies to general Hill equations,
we use 1D layered photonic crystals as a concrete and physically transparent illustration,
since they lead directly to a Hill equation with real periodic coefficients
and admit an exact transfer-matrix description.
This setting allows all aspects of the Floquet--Bloch construction,
including monodromy, spectral transitions, and hybrid (Jordan) modes,
to be demonstrated without additional numerical or model-specific complications.

\section{Floquet Theorem and Monodromy Matrix}

We recall the standard construction only to fix notation.
If $E(z)$ is a solution of Eq.~(2), then periodicity implies that $E(z+d)$ is also a solution.
Let $\tilde{E}(z)$ be an arbitrary fundamental matrix of Eq.~(2),
\begin{equation}
\tilde{E}(z) = 
\left[\begin{array}{lr}
E_1(z) & E_2(z) \\
E'_1(z) & E'_2(z)
\end{array}\right],
\end{equation}
where $E_{1,2}(z)$ are linearly independent solutions.
Since $\tilde{E}(z+d)$ is again a fundamental matrix, it must be related to $\tilde{E}(z)$ 
by right multiplication by a constant invertible matrix,
\begin{equation}
\tilde{E}(z+d) = \tilde{E}(z)\,\tilde{A},
\end{equation}
where $\tilde{A}$ is the monodromy matrix,
\begin{equation}
\tilde{A} = \tilde{E}^{-1}(0)\tilde{E}(d).
\end{equation}
Because the Wronskian of Eq.~(2) is constant, $\det(\tilde{A}) = 1$.
Each choice of fundamental system generates a monodromy matrix through the definition above.
Different choices produce conjugate matrices associated with the same differential equation
and therefore share the same spectrum.
In this sense the monodromy matrix represents the one-period propagation operator
expressed in the basis determined by the chosen fundamental system.

The monodromy matrix $\tilde{A}$ admits either a diagonal or a Jordan canonical form.
Accordingly, there exists an invertible matrix $\tilde{B}$ such that either
\begin{equation}
\tilde{B}^{-1}\,\tilde{A}\,\tilde{B} = \tilde{D}, \quad 
\tilde{D} = \left[\begin{array}{lr}
\rho_1 & 0 \\
0 & \rho_2
\end{array}\right], 
\end{equation}
or
\begin{equation}
\tilde{B}^{-1}\,\tilde{A}\,\tilde{B} = \tilde{J}, \quad 
\tilde{J} = \left[\begin{array}{lr}
\rho & 1 \\
0 & \rho 
\end{array}\right].
\end{equation}
Defining the transformed fundamental matrix
\begin{equation}
\tilde{F}(z) = \tilde{E}(z)\,\tilde{B},
\end{equation}
we obtain
\begin{equation}
\tilde{F}(z+d) = \tilde{F}(z)\,\tilde{B}^{-1}\tilde{A}\tilde{B}.
\end{equation}
Hence the Floquet--Bloch fundamental matrix satisfies either
\begin{equation}
\tilde{F}(z+d) = \tilde{F}(z)\,\tilde{D},
\end{equation}
or
\begin{equation}
\tilde{F}(z+d) = \tilde{F}(z)\,\tilde{J},
\end{equation}
depending on whether $\tilde{A}$ is diagonalizable or non-diagonalizable.

At $z=0$, the columns of $\tilde{F}(0)=\tilde{E}(0)\tilde{B}$ provide initial data adapted to this canonical form:
an eigenvector pair of $\tilde{A}$ in the diagonalizable case,
and an eigenvector/generalized-eigenvector pair in the Jordan case.
Consequently, the corresponding states satisfy
\[
\begin{aligned}
F_1(z+d) &= \rho_1 F_1(z), & \quad F_2(z+d) &= \rho_2 F_2(z), \\
F_1^{\prime}(z+d) &= \rho_1 F_1^{\prime}(z), & \quad 
F_2^{\prime}(z+d) &= \rho_2 F_2^{\prime}(z),
\end{aligned}
\]
in the diagonalizable case, whereas in the Jordan case,
\[
\begin{aligned}
F_1(z+d) &= \rho F_1(z), & \quad 
F_2(z+d) &= \rho F_2(z) + F_1(z), \\
F_1^{\prime}(z+d) &= \rho F_1^{\prime}(z), & \quad 
F_2^{\prime}(z+d) &= \rho F_2^{\prime}(z) + F_1^{\prime}(z).
\end{aligned}
\]

The Floquet multipliers $\rho_{1,2}$ are the eigenvalues of $\tilde{A}$ and satisfy
$\det(\tilde{A}-\rho \tilde{I})=0$, where $\tilde{I}$ denotes the identity matrix.
Since monodromy matrices obtained from different fundamental systems are conjugate (see Section~\ref{sec:change}),
the multipliers are basis-independent.
Using $\det(\tilde{A})=1$, the characteristic equation takes the form
\begin{equation}
\rho^2 - \mathrm{tr}(\tilde{A})\rho + 1 = 0,
\end{equation}
or equivalently,
\begin{equation}
\begin{aligned}
\rho_1+\rho_2 &= a_{11}+a_{22},\\
\rho_1\rho_2 &= a_{11}a_{22}-a_{12}a_{21}=1.
\end{aligned}
\end{equation}
Introducing the Bloch wavenumber $\mu$ via $\rho_{1,2}=e^{\pm i\mu d}$ (with $\mu$ possibly complex) gives
\begin{equation}
\cos(\mu d) = \frac{a_{11}+a_{22}}{2}.
\end{equation}

The quantity $\cos(\mu d)=\tfrac12\mathrm{tr}(\tilde A)$ provides the standard spectral classification of Eq.~(2).
If $|\cos(\mu d)|<1$, the Bloch wavenumber $\mu$ is real (allowed bands).
If $|\cos(\mu d)|>1$, the system lies in a bandgap.
In this case $\mu$ is complex and may be written as
\[
\mu = \frac{N\pi}{d} + i\kappa, \quad \kappa > 0,
\]
with integer $N$ labeling the gap.
The Floquet multipliers are real and reciprocal to each other; more precisely,
\[
\rho_{1,2} = \{ e^{\kappa d},\, e^{-\kappa d} \}
\]
or
\[
\rho_{1,2} = \{ -e^{\kappa d},\, -e^{-\kappa d} \},
\]
depending on whether $N$ is even or odd.
Accordingly, one Floquet--Bloch solution decays exponentially while the other grows.
The growing solution is typically excluded in infinite or semi-infinite settings under bounded-field (or bounded-energy) conditions.

If $|\cos(\mu d)|=1$, the system is at a band edge.
At band edges the Floquet multiplier satisfies $\rho=\pm 1$;
the case $\rho=1$ corresponds to $d$-periodic solutions,
while $\rho=-1$ corresponds to antiperiodic solutions with period $2d$.
Band edges require separate consideration, 
since at a generic band edge the second Floquet--Bloch state 
acquires a hybrid (Jordan) structure, as discussed in the next section.
In the special case of an incipient band (vanishing gap), see Refs.~\cite{Morozov2011-3,Mogilner1992,Morozov2018},
both solutions may be chosen as Floquet--Bloch waves with $\rho_1=\rho_2=\pm 1$.

Thus the Floquet--Bloch structure is encoded in the spectrum and canonical form of the monodromy matrix $\tilde{A}$:
its eigenvalues determine the spectral regime, while its canonical (diagonal or Jordan) form fixes the structure of the associated states.
For real-valued coefficients $k^2 n^2(z)$, further structure can be made explicit, as summarized below.

\medskip

\textbf{Real Floquet--Bloch basis for real periodic coefficients.}

When the coefficient $k^2 n^2(z)$ is real-valued, one may choose a real fundamental matrix,
in which case the corresponding monodromy matrix $\tilde{A}$ is real.
Its spectral type is determined by $\mathrm{tr}(\tilde{A})$,
leading to the standard elliptic, hyperbolic, and parabolic classification.
Since $\mathrm{tr}(\tilde{A})=2\cos(\mu d)$, these regimes correspond to
$|\mathrm{tr}(\tilde{A})|>2$, $|\mathrm{tr}(\tilde{A})|<2$, and $|\mathrm{tr}(\tilde{A})|=2$, respectively.

In the hyperbolic regime, $|\mathrm{tr}(\tilde{A})|>2$ (bandgaps),
the multipliers $\rho_{1,2}$ are real and reciprocal to each other,
and the Floquet--Bloch states may be chosen real-valued.

In the parabolic case, $|\mathrm{tr}(\tilde{A})|=2$ (band edges),
the multiplier is double and equal to $\rho=\pm 1$.
A real periodic (or antiperiodic) Floquet--Bloch wave may be selected,
and in the non-diagonalizable case a real hybrid Floquet mode may likewise be constructed.

In the elliptic regime, $|\mathrm{tr}(\tilde{A})|<2$ (allowed bands),
the multipliers form a complex-conjugate pair $\rho_{1,2}=e^{\pm i\mu d}$.
The corresponding eigenvectors of the real matrix $\tilde{A}$ may be chosen
as a complex-conjugate pair, so the Floquet--Bloch waves $F_{1,2}(z)$ form a conjugate pair.
A real fundamental system may then be obtained by taking the real and imaginary parts of one Floquet--Bloch wave.
In this real basis, the monodromy over one period takes the form
\begin{equation*}
\begin{bmatrix}
F_1(z+d) \\
F_2(z+d)
\end{bmatrix}
=
\begin{bmatrix}
\cos(\mu d) & -\sin(\mu d) \\
\sin(\mu d) & \cos(\mu d)
\end{bmatrix}
\begin{bmatrix}
F_1(z) \\
F_2(z)
\end{bmatrix}.
\end{equation*}
Thus in allowed bands the monodromy acts as a rotation through angle $\mu d$
in the plane of real Floquet--Bloch solutions.

Accordingly, although the construction presented below naturally produces
complex Floquet--Bloch waves in the elliptic regime,
a purely real representation is always available for real coefficients.
The distinction between real and complex Bloch representations
is therefore a matter of basis choice and does not affect
the intrinsic spectral information encoded in $\tilde{A}$.

\section{Construction of Floquet--Bloch States}

While Section~2 establishes the existence of a Floquet--Bloch basis
through the eigenstructure of the monodromy matrix,
it does not provide an explicit formula for the transformation
that maps a given fundamental system $\tilde{E}(z)$
to the corresponding Floquet--Bloch basis $\tilde{F}(z)$.
In classical presentations of Floquet theory, this reduction is usually
stated abstractly via diagonalization or Jordan decomposition of the
monodromy matrix $\tilde{A}$.

In this section we make this reduction fully explicit.
The formulas derived below express the Floquet--Bloch states directly
in terms of the entries of the monodromy matrix associated with an
arbitrary fundamental system, thereby turning the transformation to
the Floquet--Bloch basis into a concrete computational procedure.

Starting from an arbitrary fundamental matrix,
we construct in closed form a matrix $\tilde B$
that brings the monodromy matrix $\tilde A$
to its canonical (diagonal or Jordan) form.
This yields explicit algebraic expressions for the
Floquet--Bloch states in terms of the original solutions,
with the different spectral cases treated separately.

\textbf{1) Bandgaps and allowed bands.}
We first treat the generic diagonalizable case, characterized by
\[
\mathrm{tr}[\tilde{A}] \equiv a_{11}+a_{22} \neq \pm 2,
\]
for which the multipliers satisfy $\rho_1 \neq \rho_2$.
An explicit diagonalizing matrix may be chosen as
\begin{equation}
\tilde{B}=
\begin{bmatrix}
1 & \displaystyle \frac{a_{12}}{\rho_2-a_{11}} \\[6pt]
\displaystyle \frac{a_{21}}{\rho_1-a_{22}} & 1
\end{bmatrix}.
\end{equation}
In this case, both Floquet--Bloch states are Floquet--Bloch waves, and in accordance with Eq.~(14) they are
\begin{equation}
\begin{aligned}
F_1(z) & = E_1(z) + \frac{a_{21}}{\rho_1-a_{22}}\,E_2(z),\\
F_2(z) & = \frac{a_{12}}{\rho_2-a_{11}}\,E_1(z) + E_2(z),
\end{aligned}
\end{equation}
where $F_{1,2}(z+d)=\rho_{1,2}\,F_{1,2}(z)$, and the multipliers $\rho_{1,2}$ follow from Eq.~(18) or Eq.~(19).
%Since $\rho_1 \neq \rho_2$, the system lies away from band edges.
These formulas express the Floquet--Bloch waves directly
as linear combinations of the original fundamental solutions $E_{1,2}(z)$,
with coefficients determined solely by the entries of $\tilde{A}$.

If one or both off-diagonal elements $a_{12}$ and $a_{21}$ vanish,
then $\tilde{A}$ becomes triangular and its eigenvalues satisfy
$\rho_{1,2}\in\{a_{11},a_{22}\}$.
In this case we set $\rho_1=a_{11}$ and $\rho_2=a_{22}$
to maintain regularity of the transformation coefficients
in the above expressions for $F_{1,2}(z)$.

\textbf{2) Band-edge case.}
When
\[
\mathrm{tr}[\tilde{A}] \equiv a_{11}+a_{22} = \pm 2,
\]
Eqs.~(18)~-~(19) immediately yield a double Floquet multiplier  
\[
\rho_1=\rho_2\equiv \rho=\tfrac12(a_{11}+a_{22})=\pm 1,
\]
and consequently $\cos(\mu d)=\pm 1$, see Eq.~(20).
The system is therefore at a band edge of the spectrum.
The construction of the Floquet--Bloch basis now depends on whether
the off-diagonal entries of $\tilde A$ vanish,
since this determines whether the matrix is diagonalizable or defective.

2a) In the band-edge case $a_{11}+a_{22}=\pm 2$ with at least one off-diagonal element nonzero,
the matrix $\tilde{A}$ is not diagonalizable and must be brought to the Jordan form of Eq.~(13).
A suitable choice of $\tilde{B}$ that brings $\tilde{A}$ to the Jordan form
\[
\tilde{B}^{-1}\tilde{A}\tilde{B}
=
\begin{bmatrix}
\rho & 1 \\
0 & \rho
\end{bmatrix}
\]
is given by the following subcases:
\begin{equation}
\tilde{B} =
\begin{cases}
\begin{bmatrix}
1 & \dfrac{a_{12}-1}{\rho-a_{11}} \\[8pt]
\dfrac{a_{21}}{\rho-a_{22}} & 1
\end{bmatrix},
& \text{(i) } a_{12}\neq 0,\; a_{21}\neq 0, \\[14pt]

\begin{bmatrix}
1 & 0 \\
0 & \dfrac{1}{a_{12}}
\end{bmatrix},
& \text{(ii) } a_{12}\neq 0,\; a_{21}=0, \\[14pt]

\begin{bmatrix}
0 & \dfrac{1}{a_{21}} \\
1 & 0
\end{bmatrix},
& \text{(iii) } a_{12}=0,\; a_{21}\neq 0.
\end{cases}
\end{equation}
The first Floquet--Bloch state is given by
\begin{widetext}
\begin{equation}
F_1(z)=
\begin{cases}
E_1(z)+\displaystyle\frac{a_{21}}{\rho-a_{22}}\,E_2(z),
& \text{(i) } a_{12}\neq 0,\; a_{21}\neq 0, \\[10pt]
E_1(z),
& \text{(ii) } a_{12}\neq 0,\; a_{21}=0, \\[10pt]
E_2(z),
& \text{(iii) } a_{12}=0,\; a_{21}\neq 0.
\end{cases}
\end{equation}
\end{widetext}
It satisfies $F_1(z+d)=\rho\,F_1(z)$ and is therefore periodic when $\rho=1$
and antiperiodic (with period $2d$) when $\rho=-1$.
These expressions provide the periodic (or antiperiodic) Floquet--Bloch wave explicitly 
as a linear combination of the original fundamental solutions,
with coefficients determined solely by the entries of $\tilde A$.
Note that when both $a_{12}$ and $a_{21}$ are nonzero, the identity
\[
\frac{a_{21}}{\rho-a_{22}}=\frac{\rho-a_{11}}{a_{12}}
\]
implies that the two Floquet--Bloch waves of Eq.~(22) become linearly dependent at the band edge, 
differing only by an overall scalar factor, 
and are therefore encompassed by the first case of Eq.~(23).

The second state, completing the fundamental system, is the hybrid Floquet mode defined by Eq.~(5). 
In accordance with Eq.~(14) and Eq.~(23), it is represented as
\begin{widetext}
\begin{equation}
\label{eq:F2_bandedge}
F_2(z)=
\begin{cases}
\displaystyle\frac{a_{12}-1}{\rho-a_{11}}\,E_1(z)+E_2(z),
& \text{(i) } a_{12}\neq 0,\; a_{21}\neq 0, \\[10pt]
\displaystyle\frac{1}{a_{12}}\,E_2(z),
& \text{(ii) } a_{12}\neq 0,\; a_{21}=0, \\[10pt]
\displaystyle\frac{1}{a_{21}}\,E_1(z),
& \text{(iii) } a_{12}=0,\; a_{21}\neq 0.
\end{cases}
\end{equation}
\end{widetext}
Thus, even in the non-diagonalizable case,
the complete Floquet--Bloch fundamental system
is obtained by explicit algebraic transformation
from the given fundamental matrix $\tilde E(z)$,
without recourse to canonical normalization.

2b) If, in addition to $a_{11}+a_{22}=\pm 2$, the off-diagonal entries vanish,
$a_{12}=a_{21}=0$, the monodromy matrix is already diagonal,
$\tilde A = \rho \tilde I$.
No further transformation is required:
the original solutions $E_{1,2}(z)$ form a Floquet--Bloch basis.
They are periodic when $\rho=1$ and antiperiodic (with period $2d$) when $\rho=-1$.
This is the already mentioned case of incipient bands (vanishing gaps).

While the reduction of the monodromy matrix to diagonal or Jordan form is standard,
the explicit transformation formulas given by Eqs.~(22),~(24), and~(25),
expressed directly in terms of the entries of the monodromy matrix,
provide a closed algebraic map from an arbitrary fundamental system to the
corresponding Floquet--Bloch basis.
Together with the classification of the residual representation freedom given
in Section~4, this yields an operational formulation in which the basis
dependence and its associated freedom are made explicit in a form suitable
for analytical and numerical implementation.

As a consistency check, we recover the classical formulation
obtained when the fundamental system is chosen in canonical normalization.
If $E_1(z)=u(z)$ and $E_2(z)=v(z)$,
then the monodromy matrix takes the form
\begin{equation}
\tilde A =
\begin{bmatrix}
u(d) & v(d) \\
u'(d) & v'(d)
\end{bmatrix},
\end{equation}
and the relations given by Eq.~(18) and Eq.~(20) reduce to the familiar formulas,
see Refs.~\cite{Stoker1950,MagnusWinkler1966,Eastham1975,Morozov2011-3,Morozov2018},
\begin{align}
\rho^2 - [u(d) + v'(d)]\rho + 1 &= 0, \\
\cos(\mu d) &= \frac{u(d) + v'(d)}{2}.
\end{align}

In practice, it is sufficient to construct the Floquet--Bloch basis
on a single period using Eqs.~(22), (24), and (25),
and then extend it to all $z$ via Eqs.~(16) and (17).
This yields the following expressions on the $(N+1)$th period ($N=0,1,2,\ldots$) of the structure, 
\begin{equation}
F_{1,2}(z + Nd) = \rho_{1,2}^N F_{1,2}(z), 
\end{equation} 
for the Floquet--Bloch waves, and 
\begin{equation}
F_2(z + Nd) = \rho^NF_2(z) + N\rho^{N-1}F_1(z),
\end{equation} 
for the hybrid Floquet mode.
Thus, once the Floquet--Bloch basis is determined on one cell,
its behavior throughout the entire structure follows algebraically.

\section{Change of Fundamental System and Invariance of Floquet--Bloch States}
\label{sec:change}

In the previous sections, the Floquet--Bloch fundamental matrix $\tilde{F}(z)$
was constructed from an arbitrarily chosen fundamental matrix $\tilde{E}(z)$.
Since any pair of linearly independent solutions of Eq.~(2) may be used to form $\tilde{E}(z)$,
it is natural to ask how the resulting Floquet--Bloch system depends on this choice.
As we show below, the intrinsic spectral data (Floquet multipliers and hence the Bloch wavenumber)
are basis-independent, whereas the corresponding Floquet--Bloch states are determined only up to
the elementary transformations permitted by the monodromy canonical form:
independent rescalings away from band edges and, at band edges,
upper-triangular mixing of the hybrid mode with the periodic (or antiperiodic) Bloch wave.

Let $\tilde{E}(z)$ and $\tilde{E}_{\mathrm{alt}}(z)$ be two fundamental matrices
constructed from two different pairs of linearly independent solutions of Eq.~(2).
Because both solve the same equation, they must be related by a constant invertible matrix:
\begin{equation}
\tilde{E}_{\mathrm{alt}}(z) = \tilde{E}(z)\,C.
\end{equation}

The corresponding monodromy matrices are
\begin{equation}
\tilde{A} = \tilde{E}^{-1}(0)\tilde{E}(d),
\quad
\tilde{A}_{\mathrm{alt}} =
\tilde{E}_{\mathrm{alt}}^{-1}(0)\tilde{E}_{\mathrm{alt}}(d).
\end{equation}
Using Eqs.~(31) and (32), we obtain
\begin{equation}
\tilde{A}_{\mathrm{alt}} = C^{-1}\tilde{A}\,C.
\end{equation}
Thus, the monodromy matrices obtained from two different fundamental systems are conjugate.
Consequently, they share the same Floquet multipliers $\rho_{1,2}$,
which are uniquely determined by the characteristic equation of $\tilde{A}$.
The Bloch wavenumber $\mu$ is therefore independent of the chosen basis of solutions.

Now let
\[
\tilde{F}(z)=\tilde{E}(z)\tilde{B},
\quad
\tilde{F}_{\rm alt}(z)=
\tilde{E}_{\rm alt}(z)\tilde{B}_{\rm alt},
\]
be the Floquet--Bloch fundamental systems constructed from
$\tilde{E}(z)$ and $\tilde{E}_{\rm alt}(z)$, respectively.
Using Eq.~(31) and the definition $\tilde F(z)=\tilde E(z)\tilde B$, we obtain
\[
\tilde{F}_{\rm alt}(z)
= \tilde{E}_{\rm alt}(z)\tilde{B}_{\rm alt}
= \tilde{E}(z)\,C\,\tilde{B}_{\rm alt}
= \tilde{F}(z)\,\tilde{B}^{-1} C \tilde{B}_{\rm alt}.
\]
Hence,
\begin{equation}
\label{eq:F_relation}
\tilde{F}_{\rm alt}(z) = \tilde{F}(z)\,\tilde{S},
\end{equation}
where
\begin{equation}
\label{eq:S_definition}
\tilde{S} = \tilde{B}^{-1} C \tilde{B}_{\rm alt}
\end{equation}
is a constant invertible matrix.
Thus, the Floquet--Bloch basis is unique up to right multiplication by a constant invertible matrix $\tilde{S}$.

\medskip

\textbf{Nondegenerate case ($\rho_1 \neq \rho_2$)}

If $\rho_1 \neq \rho_2$, both Floquet--Bloch matrices satisfy the monodromy 
relation given by Eq.~(16), i.e.
\[
\tilde{F}(z+d) = \tilde{F}(z)\,\tilde{D},
\quad
\tilde{F}_{\rm alt}(z+d) = \tilde{F}_{\rm alt}(z)\,\tilde{D}.
\]
Using Eq.~(34), this implies
\[
\tilde{F}(z)\,\tilde{D}\,\tilde{S}
=
\tilde{F}(z)\,\tilde{S}\,\tilde{D}.
\]
Since $\tilde{F}(z)$ is a fundamental matrix and therefore invertible for all $z$, 
this reduces to the commutation relation
\[
\tilde{D}\tilde{S}=\tilde{S}\tilde{D}.
\]

Because the diagonal entries $\rho_1$ and $\rho_2$ are distinct,
the only matrices that commute with $\tilde{D}$ are diagonal matrices, i.e.
\begin{equation}
\tilde{S}=
\begin{bmatrix}
\alpha_1 & 0 \\[3pt]
0 & \alpha_2
\end{bmatrix},
\quad \alpha_{1,2}\neq 0,
\end{equation}
and the relation $\tilde{F}_{\rm alt}(z) = \tilde{F}(z)\,\tilde{S}$ leads to the following relations between the states 
in two different Floquet--Bloch fundamental matrices
\begin{equation}
F_{1}^{(\rm alt)}(z) = \alpha_1 F_1(z), 
\quad
F_{2}^{(\rm alt)}(z) = \alpha_2 F_2(z),
\end{equation}
with constant nonzero coefficients $\alpha_1$, $\alpha_2$. 
Thus, each Floquet--Bloch wave is unique up to an overall multiplicative constant:
changing the initial solution pair merely rescales the two waves independently.
If one wishes explicit values for $\alpha_1$ and $\alpha_2$, one may fix a
normalization at some point $z_0$ (for example at $z_0=0$) and take
\begin{equation}
\alpha_1 = \frac{F_{1}^{(\rm alt)}(z_0)}{F_1(z_0)},
\quad
\alpha_2 = \frac{F_{2}^{(\rm alt)}(z_0)}{F_2(z_0)},
\end{equation}
which are independent of the choice of $z_0$,
since two Floquet--Bloch waves associated with the same multiplier can differ only by a constant factor.

\textbf{Degenerate case ($\rho_1 = \rho_2 = \rho$)}

In the generic band-edge case (non-diagonalizable $\tilde{A}$),
the second Floquet--Bloch state is necessarily a hybrid mode.
Only in the exceptional incipient-band case ($a_{12}=a_{21}=0$)
is $\tilde{A}$ diagonal with $\rho=\pm 1$,
so that both states may be chosen as periodic (or antiperiodic) Floquet--Bloch waves.

Since in the degenerate case both Floquet--Bloch matrices satisfy
\[
\tilde{F}(z+d)=\tilde{F}(z)\tilde{J},
\quad
\tilde{F}_{\rm alt}(z+d)=\tilde{F}_{\rm alt}(z)\tilde{J},
\]
substituting Eq.~(34) into the second relation yields the commutation condition,
\begin{equation}
\tilde{J}\tilde{S}=\tilde{S}\tilde{J}.
\end{equation}
It is well known that the matrices commuting with a Jordan block
are precisely the upper-triangular matrices with equal diagonal entries.
Thus, the commutation relation implies
\begin{equation}
\tilde{S}=
\begin{bmatrix}
\alpha & \beta \\[3pt]
0 & \alpha
\end{bmatrix},
\quad \alpha \neq 0.
\end{equation}
It then follows from the relation $\tilde{F}_{\rm alt}(z)=\tilde{F}(z)\tilde{S}$ for the Floquet--Bloch states that
\begin{equation}
F_{1}^{(\rm alt)}(z) = \alpha\,F_1(z), \,\,\,
F_{2}^{(\rm alt)}(z) = \alpha\,F_2(z)+\beta\,F_1(z).
\end{equation}
Therefore, at a generic band edge (Jordan case):
1) the periodic (or antiperiodic) Floquet--Bloch wave is unique up to a nonzero multiplicative constant;
2) the hybrid Floquet mode is unique up to addition of a constant multiple of the periodic (or antiperiodic) wave, 
together with the same overall scaling.

If explicit parameters are desired, they may be extracted at any point $z_0$ (for example at $z_0=0$), 
chosen such that $F_1(z_0)\neq 0$ (which is always possible since a nontrivial solution cannot vanish identically):
\begin{equation}
\alpha = \frac{F_{1}^{(\rm alt)}(z_0)}{F_1(z_0)}, \quad
\beta = \frac{F_{2}^{(\rm alt)}(z_0)-\alpha\,F_2(z_0)}{F_1(z_0)}.
\end{equation}

\textbf{Summary}

Together, Eqs.~(36) and (40) completely characterize the commutant of the canonical monodromy form 
and therefore exhaust the representation freedom of the Floquet--Bloch basis.

For any two choices of linearly independent solution pairs of Eq.~(2),
the corresponding Floquet--Bloch systems are related by a constant invertible matrix, see Eq.~(34).
The intrinsic spectral content of the Floquet--Bloch construction
is therefore independent of the choice of the fundamental solutions.
Away from band edges, this amounts only to rescaling of the
Floquet--Bloch waves, whereas at band edges the hybrid mode may change
by addition of a constant multiple of the periodic (or antiperiodic) Floquet--Bloch wave.

\section{Floquet--Bloch States and the $\tilde{W}$-Matrix}

In many physical applications, in particular in the theory of photonic crystals,
wave propagation through a periodic structure is described using a transfer matrix 
that relates the field and its derivative at two spatial points. 
For the Hill equation~(2), the natural transfer object is the $\tilde{W}$-matrix.
The definition and properties of the $\tilde{W}$-matrix can be found in Refs.~\cite{Sprung1993,Lekner1994,Morozov2011-2,Morozov2023}.
The transfer-matrix formulation is particularly convenient in the present context 
because it does not rely on a specific normalization of the underlying solutions: 
the propagation operator $\tilde W(z_2,z_1)$ is intrinsic to Eq.~(2) 
and therefore allows Floquet--Bloch states to be constructed starting from any fundamental system.
In this section we show how this property simplifies the construction of Floquet--Bloch states.

The $\tilde{W}$-matrix relates the values of any fundamental matrix $\tilde{E}(z)$ 
at two arbitrary points $z_1$ and $z_2$ along the $z$-axis,
\begin{equation}
\tilde{E}(z_{2}) = \tilde{W}(z_{2}, z_{1})\tilde{E}(z_{1}).
\end{equation}
Its explicit expression in terms of any two linearly independent solutions of Eq.~(2) is
\begin{align}
\tilde{W}(z_2,z_1)&  = \tilde E(z_{2})\,\tilde E^{-1}(z_{1}) \nonumber \\
& = \begin{bmatrix}
E_1(z_2) & E_2(z_2)\\
E_1'(z_2) & E_2'(z_2)
\end{bmatrix}
\frac{1}{w}
\begin{bmatrix}
E_2'(z_1) & -E_2(z_1)\\
-E_1'(z_1) & E_1(z_1)
\end{bmatrix},
\end{align}
where $w$ is the (constant) Wronskian.
The $\tilde{W}$-matrix does not depend on the particular choice of fundamental solutions of Eq.~(2). 
If $\tilde E_{\rm new}(z)=\tilde E(z)C$ with constant invertible $C$, then
$\tilde W_{\rm new}(z_2,z_1)=\tilde E(z_2)C(C^{-1}\tilde E^{-1}(z_1))=\tilde W(z_2,z_1)$,
so $\tilde{W}$ is independent of the chosen fundamental system.
Thus $\tilde{W}(z_2,z_1)$ is an intrinsic object associated with Eq.~(2): 
it depends only on the differential equation and the points $z_1$ and $z_2$.

The spectral role of the one-period transfer matrix, $\tilde{W}_d\equiv\tilde{W}(d,0)$, 
follows directly from the defining Floquet relation. If $F(z)$ is a Floquet--Bloch wave with multiplier
$\rho$, then
\[
\begin{bmatrix}
F(d)\\
F'(d)
\end{bmatrix}
=
\rho
\begin{bmatrix}
F(0)\\
F'(0)
\end{bmatrix}.
\]
On the other hand, by definition of the transfer matrix,
\[
\begin{bmatrix}
F(d)\\
F'(d)
\end{bmatrix}
=
\tilde{W}_d
\begin{bmatrix}
F(0)\\
F'(0)
\end{bmatrix}.
\]
Hence the initial vector
\[
\begin{bmatrix} 
F(0) \\ F'(0) 
\end{bmatrix}
\]
is an eigenvector of $\tilde W_d$ with eigenvalue $\rho$.
At band edges, where the second Floquet--Bloch state takes the hybrid (Jordan) form, 
the corresponding initial data form an eigenvector/generalized-eigenvector pair of $\tilde W_d$.  
Thus the Floquet--Bloch basis is determined by the eigenstructure of the one-period transfer matrix.

Once the Floquet--Bloch initial vectors have been determined, 
the Floquet--Bloch fundamental matrix may be propagated throughout the structure using the transfer matrix,
\begin{equation}
\tilde{F}(z) = \tilde{W}(z,0)\tilde{F}(0) = \tilde{W}(z,0)\tilde{E}(0)\tilde{B}.
\end{equation}
Thus, instead of constructing the Floquet--Bloch states throughout the
entire periodic structure [see Eqs.~(22), (24), and (25)],
one may determine them at $z=0$ and use Eq.~(45) to obtain their values at any other point.

It is also important to emphasize that $\tilde E(0)$ may be chosen freely:
any invertible $2\times 2$ matrix may be prescribed as initial data.
Once $\tilde W(z,0)$ is known, the corresponding fundamental matrix
follows as $\tilde E(z)=\tilde W(z,0)\tilde E(0)$, so the underlying
solutions need not be constructed explicitly.
This again reflects the intrinsic character of the propagation operator $\tilde W(z,0)$ 
associated with Eq.~(2) and therefore allows Floquet--Bloch states 
to be constructed from any convenient fundamental system.

Finally, it is instructive to relate the transfer matrix over one period, $\tilde{W}_d$,
to the monodromy matrix $\tilde{A}$ associated with a chosen fundamental matrix $\tilde{E}(z)$.
Using Eq.~(11), one may write
\begin{equation}
\tilde{A}=\tilde{E}^{-1}(0)\tilde{E}(d)=\tilde{E}^{-1}(0)\tilde{W}_d\tilde{E}(0).
\end{equation}
Thus $\tilde{A}$ and $\tilde{W}_d$ are conjugate matrices.
In this sense the monodromy matrix represents the one-period propagation operator
expressed in the basis determined by the chosen fundamental system.

Because conjugate matrices have identical spectra,
the Floquet multipliers $\rho_{1,2}$ coincide with the eigenvalues of $\tilde{W}_d$.
Consequently, the full spectral classification of Floquet--Bloch states
(allowed bands, bandgaps, band edges, including incipient bands)
may be obtained directly from the transfer matrix $\tilde{W}_d$,
equivalently to the standard monodromy formulation.
This observation further confirms that the Floquet--Bloch construction
presented in this work may be carried out directly in any convenient solution basis.

\section{Practical Recipe}
\label{sec:recipe}

We now summarize the preceding analysis as a step-by-step procedure 
for constructing Floquet--Bloch states from arbitrary fundamental systems of Eq.~(2). 
Two equivalent implementations are given: a direct construction over one period, 
and a transfer-matrix-based construction using $\tilde{W}$. 
In both cases, the spectral data are basis-independent. 
Different initial choices affect the resulting states 
only through the representation freedoms classified in Section~\ref{sec:change}.

\medskip

\noindent\textbf{Direct construction (without transfer matrix).}

\begin{enumerate}
\item \textbf{Compute a fundamental system on the first period.}  
Compute any pair of linearly independent solutions $E_1(z)$ and $E_2(z)$
of Eq.~(2) on the interval $0 \le z \le d$.

\item \textbf{Form the monodromy matrix.}  
Construct the fundamental matrix $\tilde{E}(z)$ from $E_1(z)$ and $E_2(z)$
and evaluate
\[
\tilde{A} = \tilde{E}^{-1}(0)\tilde{E}(d),
\]
see Eq.~(11).

\item \textbf{Construct Floquet--Bloch states on the first period.}  
\begin{itemize}
\item If $\mathrm{tr}[\tilde{A}] \neq \pm 2$ (allowed bands or bandgaps),
use Eq.~(22) to construct the two Floquet--Bloch waves.

\item If $\mathrm{tr}[\tilde{A}] = \pm 2$ (band edges), distinguish:
\begin{itemize}
\item[(a)] If at least one off-diagonal element of $\tilde{A}$ is nonzero (generic band edge),
use Eqs.~(24) and (25) to construct the Floquet--Bloch wave and the hybrid Floquet mode.
\item[(b)] If $a_{12}=a_{21}=0$ (incipient band, $\tilde{A}=\rho\tilde I$),
the solutions $E_1(z)$ and $E_2(z)$ are already two independent periodic ($\rho=1$)
or antiperiodic ($\rho=-1$) Floquet--Bloch waves.
\end{itemize}
\end{itemize}

\item \textbf{Extend the states to all periods.}  
For Floquet--Bloch waves, use Eq.~(29).
For the hybrid Floquet mode, use Eq.~(30).
In incipient bands, the Floquet--Bloch states are already periodic or antiperiodic,
so extension is immediate.
\end{enumerate}

\medskip

\noindent\textbf{Optimized construction (with transfer matrix).}

Assume that the transfer matrix $\tilde{W}(z,0)$ is known on the first period $0 \le z \le d$.

\begin{enumerate}
\item \textbf{Specify a fundamental matrix at the origin.}  
Choose any invertible $2 \times 2$ matrix as $\tilde{E}(0)$.
As discussed in Section~5, this uniquely specifies a fundamental system
of solutions of Eq.~(2) via its initial data.

\item \textbf{Form the monodromy matrix.}  
Compute
\[
\tilde{A} = \tilde{E}^{-1}(0)\,\tilde{W}_d\,\tilde{E}(0),
\]
see Eq.~(46).

\item \textbf{Construct Floquet--Bloch states at the origin.}  
Proceed exactly as in Step~3 of the direct construction above,
using the monodromy matrix $\tilde A$ just computed,
distinguishing between nondegenerate bands, generic band edges,
and incipient bands (case~2b of Section~3),
but apply the construction only to determine the Floquet--Bloch initial vectors $F_1(0)$ and $F_2(0)$,
rather than the functions $F_1(z)$ and $F_2(z)$ defined on $0 \le z \le d$.

\item \textbf{Propagate across the first period.}  
For $0 \le z \le d$, evaluate
\[
\tilde{F}(z) = \tilde{W}(z,0)\tilde{F}(0).
\]

\item \textbf{Extend to all subsequent periods.}  
Use Eq.~(29) for Floquet--Bloch waves and Eq.~(30) for the hybrid Floquet mode.
In incipient bands, periodicity or antiperiodicity makes further extension trivial.
\end{enumerate}

\medskip

\noindent\textbf{Comment.}  
The transfer-matrix implementation avoids computing individual solutions
$E_1(z)$ and $E_2(z)$ throughout the first period.
Floquet--Bloch states are constructed at a single reference point
($z=0$) and propagated using $\tilde{W}(z,0)$.
This approach is typically more efficient and numerically robust,
particularly near band edges where direct construction may suffer from
loss of linear independence.

\section{Example: Multilayered Crystal}

We now illustrate the general results from the previous sections
using a photonic crystal whose period consists of two homogeneous dielectric layers 
with refractive indices $n_j$ and thicknesses $d_j$ ($j = 1,2$), see Fig.~1.
The corresponding wavenumbers in each layer are $k_j = k n_j$, and the overall period is $d = d_1 + d_2$.
Such a piecewise-constant refractive index profile leads to the Hill equation~(2),
which is closely related to the classical Meissner equation \cite{Meissner1918}.
Here, however, we analyze it using the transfer-matrix formulation,
which is naturally suited to wave propagation in layered media.

\begin{figure}[htb]
\centering
\includegraphics[width=0.95\columnwidth]{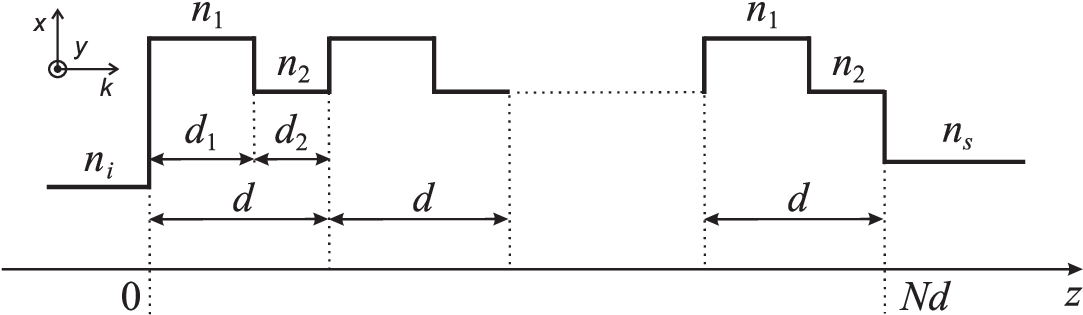}
\caption{Schematic representation of a binary photonic crystal, deposited on a substrate with refractive index $n_s$.
The light of vacuum wavenumber $k$ impinges on the crystal parallel to its axis $z$.}
\end{figure}

\subsection{Transfer matrix for a binary period}

Following Ref.~\cite{Morozov2023}, the transfer matrix $\tilde{W}(z,0)$
within the first period $0 \le z \le d = d_1 + d_2$ is given by
\begin{equation}
\tilde{W}(z,0) =
\begin{bmatrix}
\cos(k_1 z) & \sin(k_1 z)/k_1 \\
- k_1 \sin(k_1 z) & \cos(k_1 z)
\end{bmatrix},
\quad 0 \le z \le d_1,
\end{equation}
and
\begin{align}
\tilde{W}(z,0) & =
\begin{bmatrix}
\cos\bigl(k_2(z-d_1)\bigr) & \sin\bigl(k_2(z-d_1)\bigr)/k_2 \\
- k_2 \sin\bigl(k_2(z-d_1)\bigr) & \cos\bigl(k_2(z-d_1)\bigr)
\end{bmatrix} \nonumber\\
& \times \begin{bmatrix}
\cos(k_1 d_1) & \sin(k_1 d_1)/k_1 \\
- k_1 \sin(k_1 d_1) & \cos(k_1 d_1)
\end{bmatrix},
\quad d_1 \le z \le d.
\end{align}

The elements of the transfer matrix $\tilde{W}_d$ are therefore
\begin{equation}
\begin{aligned}
[\tilde{W}_d]_{11} & = \cos(k_2 d_2)\cos(k_1 d_1)
      - \frac{k_1}{k_2}\sin(k_2 d_2)\sin(k_1 d_1), \\[2pt]
[\tilde{W}_d]_{21} & = -k_2 \sin(k_2 d_2)\cos(k_1 d_1)
                 - k_1 \cos(k_2 d_2)\sin(k_1 d_1),\\[2pt]
[\tilde{W}_d]_{12} & = \frac{1}{k_2}\sin(k_2 d_2)\cos(k_1 d_1)
       + \frac{1}{k_1}\cos(k_2 d_2)\sin(k_1 d_1), \\[2pt]
[\tilde{W}_d]_{22} & = \cos(k_2 d_2)\cos(k_1 d_1)
                 - \frac{k_2}{k_1}\sin(k_2 d_2)\sin(k_1 d_1). 
\end{aligned}
\end{equation}

\subsection{Choice A: identity fundamental matrix}

We begin by choosing an arbitrary invertible matrix at $z=0$, which serves as initial data
for the transfer-matrix propagation.
Propagation by $\tilde{W}(z,0)$ then uniquely determines a fundamental matrix
$\tilde{E}(z)=\tilde{W}(z,0)\tilde{E}(0)$ of Eq.~(2) for $z\ge 0$, 
since $\tilde{W}(z,0)$ is the transfer matrix associated with Eq.~(2).

For concreteness, we first take
\begin{equation}
\tilde{E}(0) = 
\begin{bmatrix}
1 & 0 \\
0 & 1
\end{bmatrix}.
\end{equation}
This choice corresponds to a fundamental matrix $\tilde{E}(z)$ in the first layer of the crystal of the form
\begin{equation*}
\tilde{E}(z) = 
\begin{bmatrix}
\cos(k_1z) & \sin(k_1z)/k_1 \\
-k_1\sin(k_1z) & \cos(k_1z)
\end{bmatrix},
\end{equation*}
i.e. it corresponds to the standard normalization $\tilde{E}(0) = \tilde{I}$
used to define the canonical pair $u(z)$ and $v(z)$ (Section~3), with $\tilde{E}(z)$ obtained by transfer-matrix propagation.

Since $\tilde{E}(0)=\tilde{I}$, Eq.~(46) implies that the monodromy matrix $\tilde{A}_{\rm n}$ coincides with $\tilde{W}_d$,
\begin{equation}
\tilde{A}_{\rm n} = \tilde{W}_d.
\end{equation}
Then, Eq.~(18) yields the Floquet multipliers
\begin{equation}
\rho_{1,2}
= \frac{\mathrm{tr}(\tilde{W}_d) \pm 
\sqrt{\mathrm{tr}^2(\tilde{W}_d) - 4}}{2}.
\end{equation}
The corresponding dispersion relation for the Bloch wavenumber $\mu$ in a binary photonic crystal then follows from Eq.~(20),
\begin{align}
\cos(\mu d) = & \cos(k_1 d_1)\cos(k_2 d_2) \nonumber\\
- & \frac{1}{2}\left(\frac{k_1}{k_2}+\frac{k_2}{k_1}\right)
\sin(k_1 d_1)\sin(k_2 d_2).
\end{align}
Equation~(23) gives the Floquet--Bloch fundamental matrix at the point $z=0$ in the allowed bands and bandgaps,
\begin{equation}
\tilde{F}_{\mathrm{n}}(0) =
\begin{bmatrix}
1 & \displaystyle \frac{[\tilde{W}_d]_{12}}{\rho_2 - [\tilde{W}_d]_{11}} \\
\displaystyle \frac{[\tilde{W}_d]_{21}}{\rho_1 - [\tilde{W}_d]_{22}} & 1
\end{bmatrix}.
\end{equation}
In the special case where one or both off-diagonal elements $[\tilde{W}_d]_{12}$ and $[\tilde{W}_d]_{21}$ vanish,
the multipliers simplify to
\[
\rho_1 = [\tilde{W}_d]_{11}, \quad
\rho_2 = [\tilde{W}_d]_{22},
\]
see the discussion following Eq.~(22). This avoids the apparent singularities in Eq.~(54).

Once the Floquet--Bloch waves have been constructed at $z=0$, 
their values throughout the first period are obtained via
\begin{equation}
\tilde{F}_{\mathrm{n}}(z)
= \tilde{W}(z,0)\,\tilde{F}_{\mathrm{n}}(0),
\quad 0\le z\le d,
\end{equation}
and are extended to all subsequent periods using Eq.~(29), since the system lies away from band edges.

\subsection{Choice B: traveling-wave fundamental matrix}

We now consider an alternative choice of the initial invertible matrix at $z=0$,
\begin{equation}
\tilde{E}_{\mathrm{pw}}(0) =
\begin{bmatrix}
1 & 1 \\
i k_1 & - i k_1
\end{bmatrix},
\quad k_1 = k n_1 .
\end{equation}
This choice corresponds to taking, in the first layer of the crystal, the traveling-wave fundamental matrix
\begin{equation*}
\tilde{E}(z) = 
\begin{bmatrix}
\exp(i k_1 z) & \exp(-i k_1 z) \\
i k_1 \exp(i k_1 z) & -i k_1\exp(-i k_1 z)
\end{bmatrix}.
\end{equation*}
Using Eq.~(46), the corresponding monodromy matrix is
\begin{equation}
\tilde{A}_{\mathrm{pw}}
= \tilde{E}_{\mathrm{pw}}^{-1}(0)\,\tilde{W}_d\,\tilde{E}_{\mathrm{pw}}(0),
\end{equation}
with the elements
\begin{equation}
\begin{aligned}
[\tilde{A}_{\mathrm{pw}}]_{11} = & \left\{\cos(k_2d_2)+\frac{i}{2}\left[\frac{k_1}{k_2}+\frac{k_2}{k_1}\right]\sin(k_2d_2)\right\}\,e^{ik_1d_1},\\
[\tilde{A}_{\mathrm{pw}}]_{21} = & \frac{1}{2}i\left[\frac{k_1}{k_2}-\frac{k_2}{k_1}\right]\sin(k_2d_2)\,e^{ik_1d_1},\\
[\tilde{A}_{\mathrm{pw}}]_{12} = & -\frac{1}{2}i\left[\frac{k_1}{k_2}-\frac{k_2}{k_1}\right]\sin(k_2d_2)\,e^{-ik_1d_1},\\
[\tilde{A}_{\mathrm{pw}}]_{22} = & \left\{\cos(k_2d_2)-\frac{i}{2}\left[\frac{k_1}{k_2}+\frac{k_2}{k_1}\right]\sin(k_2d_2)\right\}\,e^{-ik_1d_1}.
\end{aligned}
\end{equation}
As expected, $\tilde{A}_{\mathrm{pw}}$ has the same trace and eigenvalues as $\tilde{A}_{\mathrm{n}} = \tilde{W}_d$. 

Equation~(23) gives the Floquet--Bloch fundamental matrix at the point $z=0$ in the allowed bands and bandgaps as
\begin{equation}
\tilde{F}_{\mathrm{pw}}(0) =
\begin{bmatrix}
1 & \displaystyle \frac{[\tilde{A}_{\mathrm{pw}}]_{12}}{\rho_2 - [\tilde{A}_{\mathrm{pw}}]_{11}} \\
\displaystyle \frac{[\tilde{A}_{\mathrm{pw}}]_{21}}{\rho_1 - [\tilde{A}_{\mathrm{pw}}]_{22}} & 1
\end{bmatrix}.
\end{equation}
The Floquet--Bloch waves throughout the first period are then obtained via
\begin{equation}
\tilde{F}_{\mathrm{pw}}(z)
= \tilde{W}(z,0)\,\tilde{F}_{\mathrm{pw}}(0),
\quad 0\le z\le d,
\end{equation}
and are extended to all subsequent periods with the aid of Eq.~(29).
These Floquet--Bloch waves differ only by independent rescalings from the Floquet--Bloch waves constructed in Choice~A,
in accordance with the results of Section~\ref{sec:change}.
This is illustrated below.

We consider infrared light impinging from air ($n_i = 1$)
onto a binary photonic crystal (PC) consisting of $N = 6$ periods, 
with layers having refractive indices $n_1 = 4.0$ (Ge) and $n_2 = 2.2$ (ZnS),
and thicknesses $d_1 = 0.55~\mu m$ and $d_2 = 1.00~\mu m$, 
deposited on a glass substrate with $n_s = 1.5$. 
This setup is within the scope of the experiment described in Ref.~\cite{Wang2014}.
The transmittance and the band structure in the case of normal incidence are shown in Fig.~2.
Due to condition $n_1d_1 = n_2d_2$, every even bandgap is suppressed, see Ref.~\cite{Morozov2011-3}, 
forming the incipient bands (vanishing gaps) discussed above, while the odd bandgaps open to their maximum widths. 

\begin{figure}[htb]
\centering
\includegraphics[width=\columnwidth]{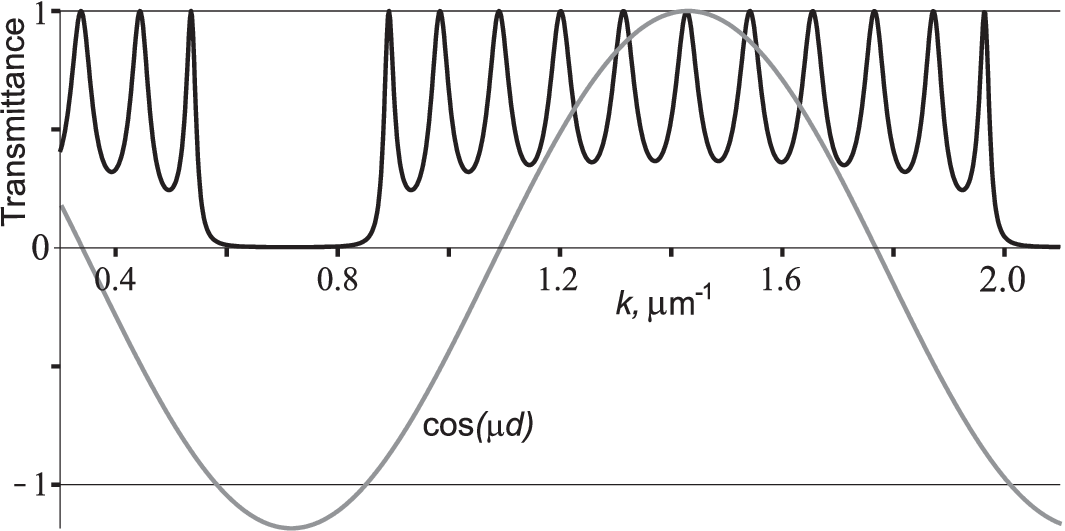}
\caption{Transmittance and $\cos(\mu d)$ for infrared light incident normally from air ($n_i = 1$) on a binary photonic crystal of $N = 6$ periods.
Each period consists of Ge ($n_1 = 4.0$, $d_1 = 0.55~\mu m$) and ZnS ($n_2 = 2.2$, $d_2 = 1.00~\mu m$) layers, 
deposited on a glass substrate with refractive index $n_s = 1.5$.
The Bloch wavenumber $\mu$ is obtained from Eq.~(52).
Allowed bands ($\lvert \cos(\mu d) \rvert  < 1$) and bandgaps ($\lvert \cos(\mu d) \rvert > 1$) are clearly observed in the transmission spectrum.
The second bandgap is suppressed due to condition $n_1d_1 = n_2d_2$.}
\end{figure}

To illustrate the invariance of the Floquet--Bloch states under different initial fundamental matrices, 
Fig.~3 and Fig.~4 show the Floquet--Bloch waves $F_{1,2}(z)$ over the first three periods of the photonic crystal, 
constructed using Choices A and B for representative wavenumbers 
$k = 0.53~\mu m^{-1}$ in an allowed band (Fig.~3) 
and $k = 0.83~\mu m^{-1}$ in a bandgap (Fig.~4).  
In full agreement with the analysis of Section~\ref{sec:change}, 
the two constructions differ only by constant ($z$-independent) complex factors,
\begin{equation}
\alpha_1 = \frac{F_1^{({\rm pw})}(z)}{F_1^{({\rm n})}(z)}, \quad 
\alpha_2 = \frac{F_2^{({\rm pw})}(z)}{F_2^{({\rm n})}(z)}.
\end{equation}

\begin{figure}[htb]
\centering
\includegraphics[width=\columnwidth]{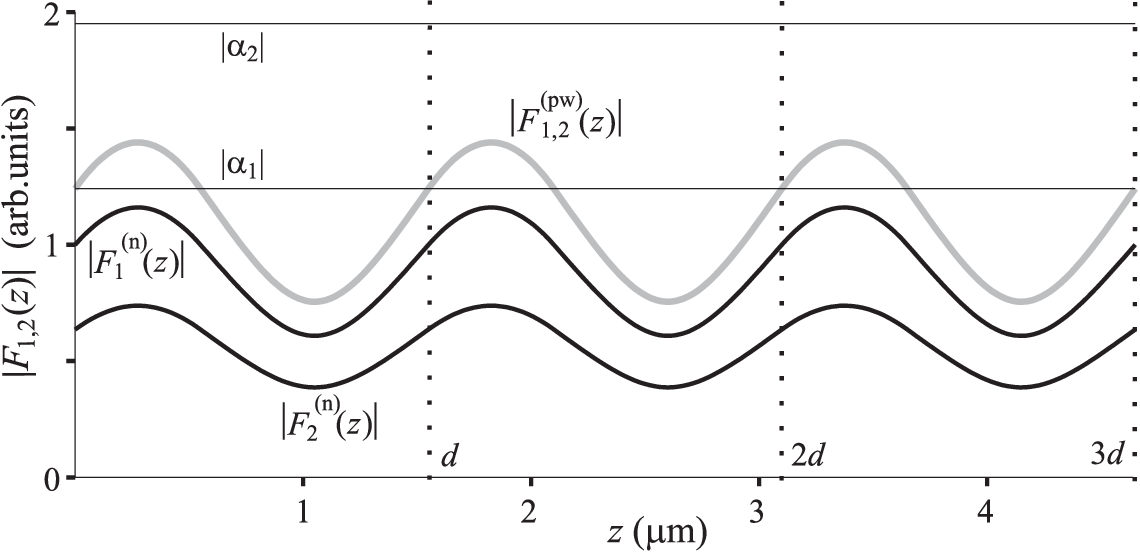}
\caption{Absolute values of the Floquet--Bloch waves over the first three periods of the binary photonic crystal, 
with parameters as in Fig.~2, for a representative wavenumber $k = 0.53~\mu m^{-1}$ in an allowed band. 
The states $F_{1,2}^{(\mathrm{n})}(z)$, shown by black lines, are constructed using Choice A (identity fundamental matrix), 
and the states $F_{1,2}^{(\mathrm{pw})}(z)$, shown by thick gray lines, are constructed using Choice B (traveling-wave fundamental matrix).
The two constructions differ only by constant complex factors $\alpha_{1,2}$, 
in full agreement with the invariance analysis of Section~\ref{sec:change}.
The states $F_{1}^{(\rm pw)}(z)$ and $F_{2}^{(\rm pw)}(z)$ form a complex-conjugate pair in allowed bands; 
as a result, their absolute values coincide. Vertical dotted lines mark unit-cell boundaries at $z = d, 2d, 3d$, with $d = d_1 + d_2$.}
\end{figure}

\begin{figure}[htb]
\centering
\includegraphics[width=\columnwidth]{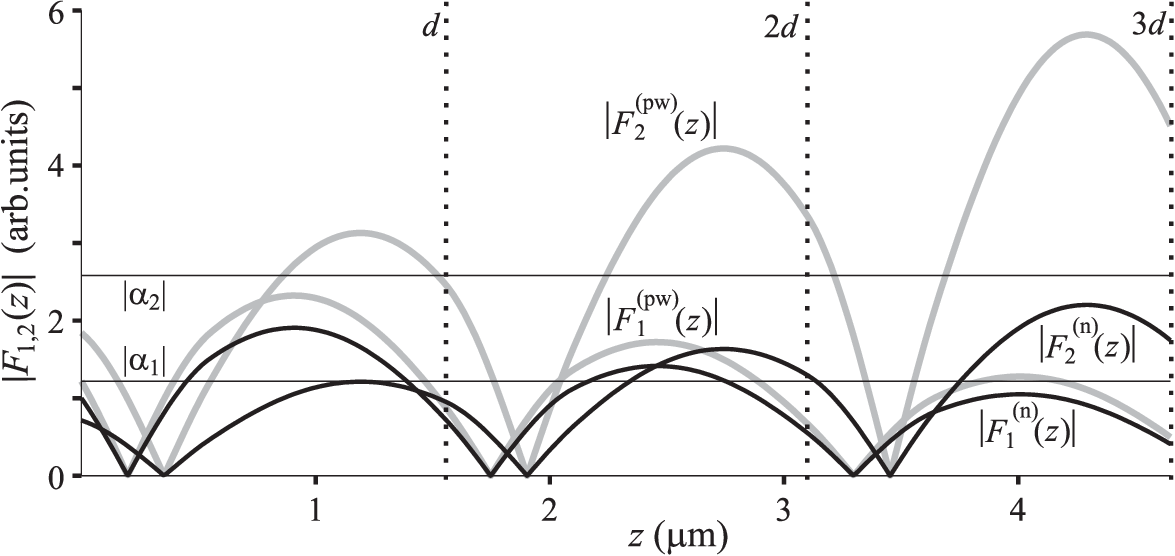}
\caption{Absolute values of the Floquet--Bloch waves over the first three periods of the binary photonic crystal, 
with parameters as in Fig.~2, for a representative wavenumber $k = 0.83~\mu m^{-1}$ in a bandgap. 
In this regime the Bloch wavenumber $\mu$ is complex, 
and one Floquet--Bloch wave is evanescent while the other grows exponentially.
As expected, the profiles obtained using Choices A and B differ only by constant complex factors $\alpha_{1,2}$,
in accordance with the invariance analysis of Section~\ref{sec:change}.}
\end{figure}

This example explicitly demonstrates that Floquet--Bloch states can be constructed from arbitrary initial fundamental systems,
without invoking normalized solutions $u(z)$ and $v(z)$.
%While many conventional presentations implicitly fix the solution basis at the beginning of the crystal,
%the present formulation shows that such a choice is inessential and amounts only to a change of representation.
By working directly with the transfer matrix and allowing complete freedom in the choice of initial data, 
the Floquet--Bloch structure emerges in a fully basis-independent manner.
This flexibility is particularly advantageous in scattering problems, layered-media optics, and numerical implementations, 
where traveling-wave formulations are often more natural and stable than
formulations based on normalized solutions $u(z)$ and $v(z)$.

\section{Conclusions}

We have developed a constructive formulation of Floquet--Bloch theory
for second-order linear differential equations with periodic coefficients,
in which the Floquet--Bloch fundamental system is obtained directly
from an arbitrarily chosen fundamental matrix.
Rather than revisiting the classical spectral results of Floquet theory,
the aim of this work is to provide an explicit, implementation-ready
formulation of the transformation to the Floquet--Bloch basis.
The resulting algebraic formulas provide closed-form expressions 
for Floquet--Bloch states in terms of the entries of the monodromy matrix 
associated with an arbitrary fundamental system, 
thereby turning the transformation to the Floquet--Bloch
basis into a concrete computational procedure.

The construction also clarifies how Floquet--Bloch states depend on the
choice of fundamental solutions.
Away from band edges the states are unique up to independent rescalings,
whereas at band edges the hybrid (Jordan) mode is defined up to the addition
of a multiple of the periodic (or antiperiodic) Bloch wave.
In this way the representation-independent content of the Floquet--Bloch
structure is made explicit.

Expressing the construction in terms of the transfer matrix
provides a convenient implementation strategy.
Although illustrated here using one-dimensional photonic crystals,
the framework applies to general Hill equations and is well suited
to analytical and numerical studies of a broad class of
one-dimensional periodic media.

%\section*{References}

%\bibliographystyle{iopart-num}
%\bibliography{photonic_crystals}

%\end{document}

\providecommand{\newblock}{}

\end{document}